\documentclass[twocolumn,english,aps,manuscript,prl,aps,reprint]{revtex4-1}
\usepackage[T1]{fontenc}
\usepackage[latin9]{inputenc}
\usepackage{amsmath}
\usepackage{amssymb}
\usepackage{graphicx}
\usepackage{esint}

\makeatletter

 \@ifundefined{textcolor}{}
 {%
   \definecolor{BLACK}{gray}{0}
   \definecolor{WHITE}{gray}{1}
   \definecolor{RED}{rgb}{1,0,0}
   \definecolor{GREEN}{rgb}{0,1,0}
   \definecolor{BLUE}{rgb}{0,0,1}
   \definecolor{CYAN}{cmyk}{1,0,0,0}
   \definecolor{MAGENTA}{cmyk}{0,1,0,0}
   \definecolor{YELLOW}{cmyk}{0,0,1,0}
 }

\usepackage{babel}

\usepackage{babel}

\usepackage{babel}

\usepackage{babel}

\makeatother

\usepackage{babel}
\begin{document}

\title{Dissipative superconductivity: a universal non-equilibrium state
of nanowires}

\author{Yu Chen$^{1,2}$, Yen-Hsiang Lin$^{2,4}$, Stephen Snyder$^{2,5}$, Allen
Goldman$^{2}$, and Alex Kamenev$^{2,3}$}

\affiliation{$^{1}$Department of Physics, University of California - Santa Barbara,
Santa Barbara, CA 93106, USA}

\affiliation{$^{2}$School of Physics and Astronomy, University of Minnesota,
Minneapolis, MN 55455, USA}

\affiliation{$^{3}$William I. Fine Theoretical Physics Institute, University
of Minnesota, Minneapolis, MN 55455, USA}

\affiliation{$^{4}$Department of  Physics, University of Michigan, Ann Arbor, MI., 48109}

\affiliation{$^{5}$ Intel Corp., Portland OR, 97124}

\date{\today}

\maketitle
\textbf{The ability to carry electric current with zero dissipation
is the hallmark of superconductivity.\cite{deGennes1999} It is this
very property which is used in applications from MRI machines to LHC
magnets. But, is it indeed the case that superconducting order is
incompatible with dissipation? One notable exception, known as vortex
flow, takes place in high magnetic fields.\cite{Blatter1994} Here
we report observation of dissipative superconductivity in far more
basic configurations: superconducting nanowires with superconducting
leads. We provide evidence that in such systems, normal current may
flow in the presence of superconducting order throughout the wire.
The phenomenon is attributed to the formation of a non-equilibrium
state, where superconductivity coexists with dissipation, mediated
by the so-called Andreev quasiparticles. Besides promise for applications
such as single-photon detectors,\cite{Goltsman2001} the effect is
a vivid example of a controllable non-equilibrium state of a quantum
liquid. Thus our findings provide an accessible generic platform to
investigate conceptual problems of out-of-equilibrium quantum systems. }

With applications ranging from infrared detectors\cite{Eisaman2011}
to prototypical qubits\cite{Clarke2008,Devoret2013}, superconducting
nanocircuitry has emerged in recent years as a fascinated area of
research. Its fundamental significance lies in\textit{ e.g}. phase-sensitive
studies of pairing mechanisms in novel superconductors\cite{Wollman1993,Tsuei1994}
and access to a wealth of non-equilibrium quantum phenomena.\cite{Mooij2006}
A key element of such circuitry - superconducting nanowires, are known
to be susceptible to strong fluctuations. Their most spectacular manifestations
are phase slips (PSs) of the superconducting order parameter, which
lead to dissipation within a nominally dissipationless superconducting
state.\cite{Bezryadin2012,Arutyunov2008,Altomare2013} Observing and
studying such dissipative superconductivity has turned out to be a
challenge. The culprits are non-equilibrium quasiparticles massively
generated by PSs. If not removed efficiently, they tend to overheat
the nanowire, driving it into the normal state. For example, thin
MoGe\cite{Sahu2009,Shah2008} and Al wires\cite{Li2011,Singh2013}
appear to be switched into the normal state by a single PS. The return
to the superconducting state requires a significant decrease of the
drive current, leading to hysteretic I-V characteristics.

In experiments reported here we overcome the excess heating by an
improved fabrication process (see Methods for details). Electrically
transparent interfaces between the wire and the leads allow for a
fast escape of non-equilibrium quasiparticles into the environment.
By choosing Zn as the growth material,\cite{Stuivinga1982} we are
able to fabricate quasi-1D wires whose length $L$ is significantly
shorter than the inelastic relaxation length $L_{in}$, yet much longer
than the coherence length $\xi\approx250$nm. These bring us to a
situation in which quasiparticles form a peculiar non-equilibrium
distribution, governed by Andreev reflections from the boundaries
with the superconducting leads. As a result, we observe a non-hysteretic
dissipative state, which still exhibits distinct superconducting features
such as a supercurrent and a sensitivity to weak magnetic fields.

\begin{figure*}
\includegraphics{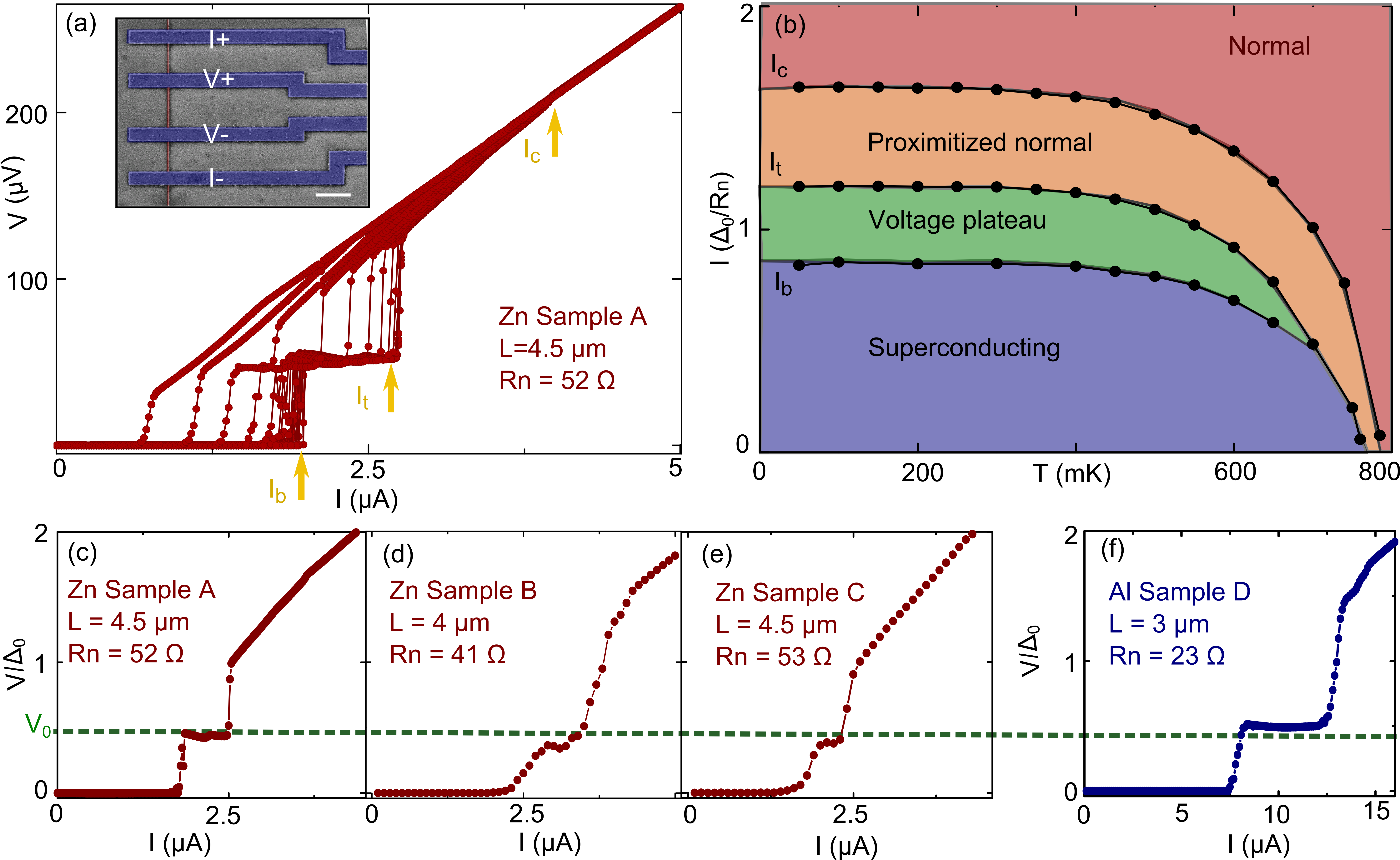}

\caption{\textbf{Experimental observation of the voltage plateau.} a) The I-V
characteristics of sample A, at temperatures from 50 mK to 750 mK
(right to left) with a 50 mK interval. The voltage plateau with $V_{0}\sim52\,\mu V$
is visible between the bottom and top threshold currents $I_{b}$
and $I_{t}$. The inset shows a scanning electron microscope image
of the sample, the scale bar represents $2\,\mu m$. b) The temperature
-- current phase diagram, showing the voltage plateau existing at
$T\lesssim650$ mK. c-f) The I-V characteristics of Zn sample A-C
and Al sample D, measured at 450 mK. Sample A was measured in the
well-filtered cryostat 1, while samples B-D -- in the weakly-filtered
cryostat 2. The voltage axes are scaled to the BCS energy gap $2\Delta_{0}=3.52T_{c}$
for each sample. The voltage plateau of all samples collapses onto
$V_{0}/\Delta_{0}\sim0.43\pm0.05$, providing the evidence for the
universality of the plateau state. }

\label{Voltage plateau}
\end{figure*}

In Fig.~\ref{Voltage plateau}a, we show I-V characteristics of sample
A, measured at temperatures from 50 to 750 mK in a well-filtered dilution
refrigerator 1 (see Methods for details). Over a range of currents,
flanked by the bottom and top threshold values $I_{b}<I<I_{t}$, the
voltage across the nanowrie exhibits a nearly flat plateau at $V_{0}=52.5\pm1.2\,\mu$V
(for $T\lesssim450$ mK). This indicates a peculiar dissipative state,
which is distinctly different from the normal state. It is important
to note that both $I_{t}$ and $I_{b}$ are factor of $30\sim50$
smaller than the estimated depairing critical current of the wire.\cite{Skocpol1974}
Collecting $I_{b}$ and $I_{t}$ together with $I_{c}$, where the
system turns normal, we construct a temperature -- current phase diagram,
shown in Fig.~\ref{Voltage plateau}b. It shows that as the temperature
increases, the voltage plateau is compressed and eventually disappears
at about 650 mK. This temperature dependence resembles that of the
superconducting order parameter, suggesting that the dissipative voltage
plateau state is associated with the superconducting order.

Performing measurements on different devices, we found a remarkable
universality associated with the voltage plateau. In Fig.~\ref{Voltage plateau}c-f,
we compare the I-V characteristics of sample A with two other Zn samples
B and C as well as an Al sample D, all measured at $450$ mK. These
samples differ from sample A both in their geometry and normal state
resistance, and were measured in a different weakly-filtered refrigerator
2. Despite some smearing by unfiltered noise, the plateau voltage
$V_{0}$ remains nearly unchanged for all of the Zn wires. The only
exception is Al sample D, with the voltage plateau at $V_{0}\simeq93.2\pm1.3\mu$V.
Rescaling this voltage with the BCS superconducting gap $2\Delta_{0}\approx3.52T_{c}$,
we found that the plateaus in all samples fall close to the same universal
line $eV_{0}/\Delta_{0}=0.43\pm0.05$ (the ratio for all samples is
listed in Supplemental Material \cite{supplemental}).

\begin{figure*}
\includegraphics{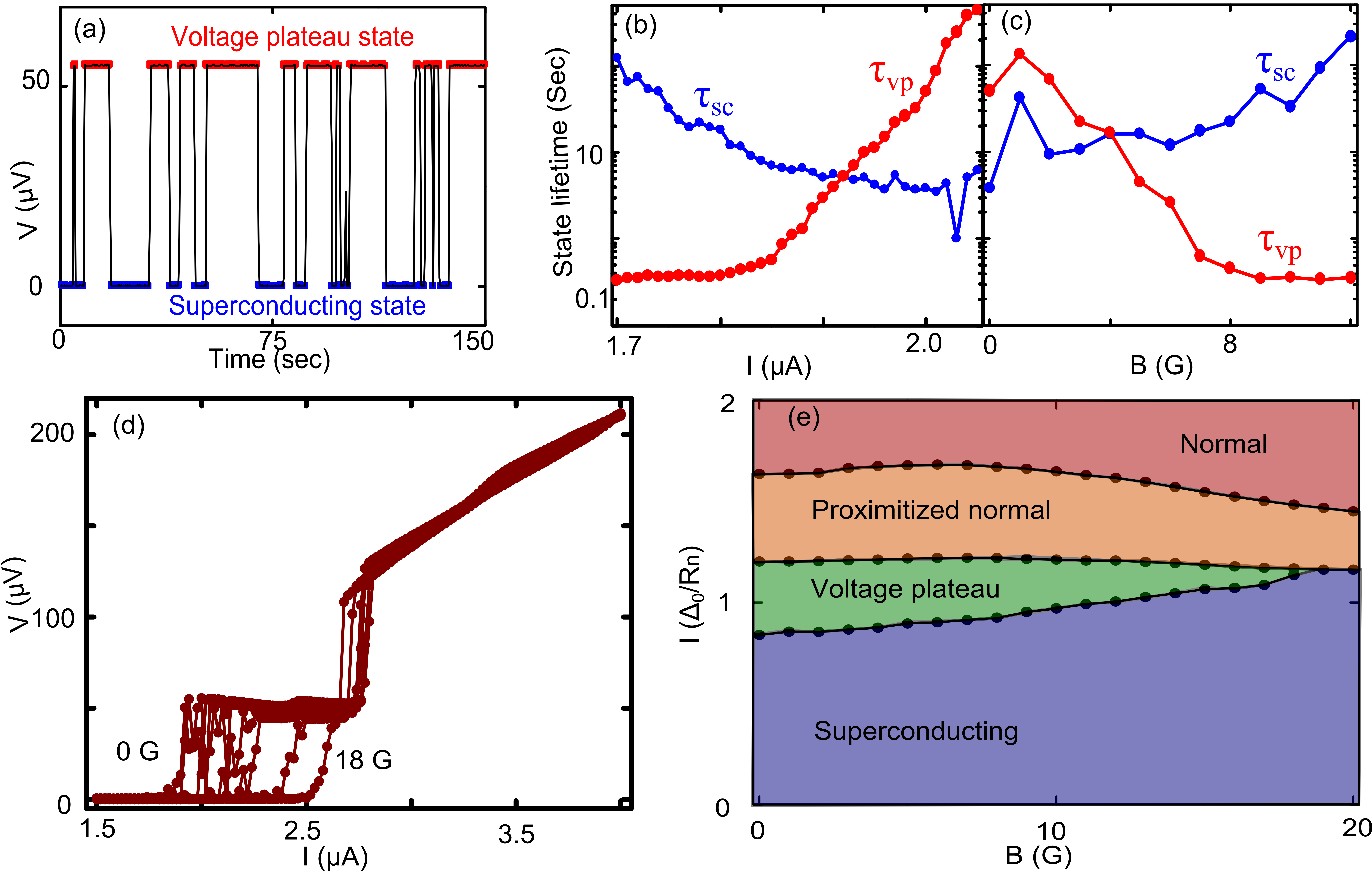}

\caption{\textbf{The onset through bistability and the magnetic response of
the voltage plateau.} a) The real-time evolution of the voltage, for
sample A hold at a fixed current of $1.95\,\mu A$ and at 50 mK. The
system undergoes stochastic switching between the superconducting
and the voltage plateau states. b) The average lifetime of the superconducting,
$\tau_{sc}$, and the voltage plateau, $\tau_{vp}$ states, at elevated
currents. The transition is accomplished by suppressing the superconducting
state and stabilizing the voltage plateau state. c) The average lifetimes
as functions of perpendicular magnetic field. A magnetic field of
only few G suppresses the lifetime of the voltage plateau state over
two orders of magnitude. d) The I-V characteristics at magnetic fields
from 0 to 18 G, with a 2 G interval. The bottom critical current $I_{b}$
is seen to increase from $\thicksim1.8\,\mu A$ to $\thicksim2.5\,\mu A$.
e) Magnetic field -- current phase diagram shows the high sensitivity
of the voltage plateau to magnetic field. The plateau disappears at
19 G.}

\label{switching}
\end{figure*}

Another remarkable feature of the voltage plateau state is its onset
through a region of stochastic bistability. It is revealed by time-domain
measurements, with the voltage measured with a repetition rate of
3 Hz under a sustained constant current. In Fig.~\ref{switching}a
we show a time trace of the measured voltage at $I=1.95\mu$A and
$T=50$mK. The system exhibits random switchings between the superconducting
and voltage plateau states with a characteristic time scale of a few
seconds, indicating an intrinsic bistability. To quantify stability
of the two competing states we define lifetimes: $\tau_{sc}$ and
$\tau_{vp}$ as the averaged residence times in the superconducting
and the voltage plateau states respectively. Figure \ref{switching}b
shows their dependencies on the applied current throughout the transition
range. Increasing the current, leads to an exponential growth of the
voltage plateau lifetime $\tau_{vp}$ and the reduction of the superconducting
lifetime $\tau_{sc}$, albeit at a smaller rate. It is worth mentioning
that the two lifetimes are nearly temperature independent until $T\sim400$
mK, above which $\tau_{vp}$ increases and $\tau_{sc}$ decreases
exponentially.\cite{supplemental}

The observed dissipative state exhibits counterintuitive magnetic
field dependence. One could expect that the magnetic field suppresses
superconductivity, thus decreasing $\tau_{sc}$ and possibly increasing
$\tau_{vs}$. In fact, the exact opposite happens. As shown in Fig.
\ref{switching}c, a magnetic field of merely a few G stabilizes the
superconducting state, {\em increasing} its lifetime by more than
an order of magnitude, simultaneously decreasing the voltage plateau
lifetime by two orders of magnitude. This behavior is consistently
observed through the entire transition range of currents.

The {\em enhancement} of superconductivity by magnetic field is
even more apparent by inspecting I-V characteristics at different
fields, Fig.~\ref{switching}d. It is evident that the field shifts
the bottom critical current $I_{b}$ to higher values, stabilizing
the superconducting state. Such a stabilization is in fact a result
of the suppression of the voltage plateau state. This is best seen
in the critical current vs. magnetic field phase diagram, Fig.~\ref{switching}e.
The range of currents supporting the voltage plateau decreases rapidly
from below until the plateau collapses at 19 G. Correspondingly the
phase space of the superconducting state expands. We thus observe
a rather counterintuitive non-equilibrium phenomenon: keeping the
current within the voltage plateau regime and {\em increasing}
the magnetic field, brings the system from the dissipative {\em
into} the superconducting state. This is consistent with the reported
\textit{magnetic-field-induced superconductivity} and \textit{anti-proximity
effect}.\cite{Chen2009,Chen2011,Tian2005,Tian2006} It is now apparent
that the magnetic field induced superconductivity originates with
the collapse of the voltage plateau state, providing an intriguing
connection between the two effects.

It is crucial to distinguish the observed voltage plateau state from
other phenomena. It is different from phase slip centers, seen in
long superconducting whiskers and characterized by a constant differential
resistance,\cite{Skocpol1974,Tidecks1990} as opposed to a constant
voltage. The plateau can't be a giant Shapiro step,\cite{Dinsmore2008,Bae2012}
caused by a leaking high frequency noise. Indeed, Zn and Al samples
have different $V_{0}$ values, requiring noise of very different
frequency within the same measurement apparatus. It also can't be
attributed to a running state of an underdamped Josephson junction.\cite{Barone1982}
The underdamped regime would require a capacitance {\em four} orders
of magnitude larger than that of our system. External capacitance
is also excluded by the fact that the same results were obtained in
two refrigerators with very different circuitry. Moreover, contrary
to the observed plateau, voltage across an underdamped junction is
expected to grow with the increased current bias.

\begin{figure*}
\includegraphics{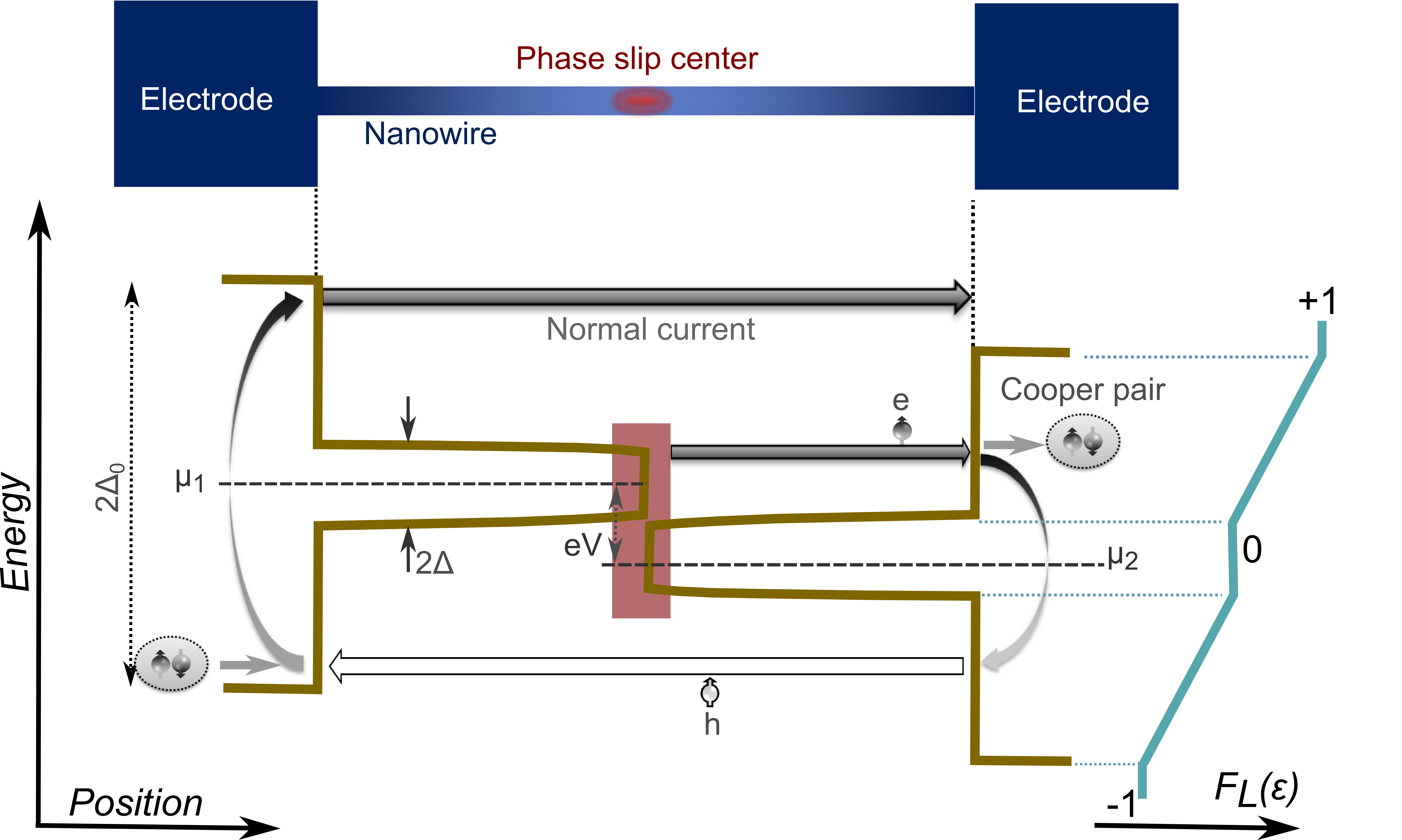}

\caption{\textbf{Dissipative superconducting state of the nanowire.} (\textit{Top})
Schematic of the system: superconducting nanowire is connected to
superconducting leads. (\textit{Bottom}) The energy gap profile as
a function of position along the system. Phase slips occur in the
center of the wire, generating non-equilibrium quasiparticles. At
the two wire-lead interfaces, they experience multiple Andreev reflections
before escaping into the leads. This establishes a universal distribution,
with the longitudinal part $F_{L}(\epsilon)$ shown on the right.
The self-consistency relation (\ref{eq:selfconsistency}) results
in the suppressed non-equilibrium gap $2\Delta\approx0.34\Delta_{0}$,
dictating that the voltage $eV_{0}=2\Delta$ should develop to maintain
PSs events.}

\label{energy}
\end{figure*}

Key to understanding these observations are non-equilibrium quasiparticles\cite{Skocpol1974}
generated by PSs. In our samples (in constrast with previous studies~\cite{Sahu2009,Shah2008,Li2011,Singh2013,Skocpol1974})
the inelastic relaxation length $L_{in}$ exceeds the wire length
\cite{Stuivinga1982}, allowing the quasiparticles to spread over
the entire wire. At the interfaces with the leads the quasiparticles
experience Andreev reflections, which mix particles and holes. This
leads to quasiparticle diffusion over energy\cite{Nagaev2001}, resulting
in a peculiar non-equilibrium distribution. Because of the self-consistency
relation, such a non-equilibrium distribution suppresses the order
parameter $\Delta$ inside the wire (relative to its equilibrium value
$\Delta_{0}$). Although the quasiparticles are far from equilibrium,
the order parameter is fixed to its local self-consistent value everywhere
apart from a distance $\sim\xi$ around the PS. The {\em condensate}
chemical potential $\mu$, given by the Josephson relation $\mu=\hbar\langle\partial_{t}\phi\rangle/2$,
thus exhibits a discontinuity $eV$ at the PS location\cite{Skocpol1974}.
Since the leads absorb high-energy quasiparticles, the concentration
of the latter is largest in the center, pinning the PS to the midpoint
of the wire. For PS's to occur, the voltage must exceed the energy
gap, Fig.~\ref{energy}, i.e. $eV_{0}\approx2\Delta$. In this case
PSs keep generating quasiparticles, self-propelling the dissipative
state.

For a quantitative description \cite{supplemental} it is convenient
to parametrize the quasiparticle distribution function $F(\epsilon,x)=1-2n(\epsilon,x)$
by its longitudinal and transverse components, which are its odd and
even parts $F_{L/T}(\epsilon,x)=F(\epsilon-\mu_{a},x)\mp F(-\epsilon+\mu_{a},x)$
with respect to the two chemical potentials $\mu_{1,2}=\pm eV/2$.
At the boundaries with the leads $x=x_{1,2}(\epsilon)$ they obey
Andreev boundary conditions:
\begin{equation}
\left.F_{T}(\epsilon,x)\right|_{x=x_{a}(\epsilon)}=0;\quad\quad\left.\partial_{x}F_{L}(\epsilon,x)\right|_{x=x_{a}(\epsilon)}=0.\label{eq:boundary}
\end{equation}
In the absence of inelastic relaxation the continuity relation (known
also as the Usadel equation\cite{Usadel1970,Kamenev2011}) reads $\partial_{x}F_{L/T}(\epsilon,x)=J_{L/T}(\epsilon)$,
which along with Eq.~(\ref{eq:boundary}) leads to $x$-independent
$F_{L}(\epsilon)$ (for $|\epsilon|<\Delta_{0}$), satisfying the
energy-diffusion equation\cite{Nagaev2001} $\partial_{\epsilon}^{2}F_{L}=0$.
Since the self-consistency relation
\begin{equation}
\int_{\Delta(x)}^{\omega_{D}}\!\! d\epsilon\,\frac{F_{L}(\epsilon,x)}{\sqrt{\epsilon^{2}-\Delta^{2}(x)}}=\int_{\Delta_{0}}^{\omega_{D}}\!\! d\epsilon\,\frac{\mathrm{tanh}(\epsilon/2T)}{\sqrt{\epsilon^{2}-\Delta_{0}^{2}}}\label{eq:selfconsistency}
\end{equation}
($\omega_{D}$ is Debye frequency) involves only $F_{L}$, it results
in an almost constant $\Delta(x)\approx\Delta$. The PSs in the middle
of the wire excite quasiparticles and holes, equilibrating their populations.
This provides the boundary condition $F_{L}(|\epsilon|<\Delta)=0$
for the energy diffusion \cite{Nagaev2001}, resulting in the distribution
function depicted in Fig.~\ref{energy}. Being substituted into the
self-consistency relation (\ref{eq:selfconsistency}), it results
in a transcendental equation for $\delta=\Delta/\Delta_{0}$, which
at $T=0$ has two solutions (bistability!): $\delta_{sc}=1$ and $\delta_{vp}=0.17$.

The second of these solutions implies a {\em dissipative} state
with $eV_{0}=2\Delta\approx0.34\Delta_{0}$. It is sustained if a
normal current $I_{b}\approx1.64(V_{0}/R_{n})$ \cite{supplemental},
is applied to the wire. Notice that for $\xi\ll L$ the current $I_{b}$
is much smaller than the deparing critical current, allowing the wire
to still support a supercurrent. An excess current $I-I_{b}$ is thus
carried as the supercurrent without an additional voltage increase
- hence the observed voltage plateau. The current exceeding $I_{t}\approx0.72\Delta_{0}/eR_{n}$
stabilizes another solution of the self-consistency and the energy
diffusion equations: the one with vanishing order parameter in the
middle of the wire. It essentially terminates the supercurrent, resulting
in the resistance being close to the normal one. The fact that these
threshold values are about 30\% less than the observed ones is attributed
to the residual inelastic processes, neglected above. Indeed, the
latter lead to particle-hole recombination, driving the distribution
towards the equilibrium one. A reasonable estimate $L_{in}\approx12\mu m$
\cite{Stuivinga1982} brings the currents $I_{b,t}$ as well as the
voltage plateau $V_{0}$ within 10\% of the observed values.

This picture also naturally accounts for the observed effects of the
magnetic field and temperature. The field mostly suppresses the order
parameter of the {\em leads} $\Delta_{lead}$, leaving that of
the wire (almost) intact. This narrows the interval for the energy
diffusion\cite{Vodolazov2012}, bringing the distribution closer to
the equilibrium one. This in turn increases the bottom threshold current
$I_{b}$. For $\Delta_{lead}\lesssim0.78\Delta_{0}$ the self-consistent
solution of Fig.~\ref{energy} with $\delta\neq1$ is not possible
anymore. The superconducting state of the wire is thus stabilized
all the way up to $I_{t}$. In fact, suppression of $\Delta_{lead}$
is also the primary mechanism of the voltage plateau termination at
$T\gtrsim650$ mK, Fig.~\ref{Voltage plateau}b.

\section{Methods}

Investigated Zn nanowires are 100--120 nm wide, 65--110 nm high, with
the length 1.5$\mu$m <L<6$\mu$m (see Table I in Supplemental material)
connected to four 1 $\mu$m wide Zn electrodes, inset to Fig.~\ref{Voltage plateau}a.
The Zn electrodes are 10 $\mu$m long and in turn are connected to
pre-patterned Au contacts. Both the nanowire and the electrodes were
fabricated in a single step of the quench-deposition at 77K substrate
temperatures, depositing through a resist mask patterned using electron-beam
lithography. The normal sate resistivity varyies $\rho=(6.2-8.4)*10^{-8}\Omega$m,
in comparison the bulk value for Zn is $\rho=5.9*10^{-8}\Omega$m.

The electrical measurements were carried out in two different refrigerators:
(1) Oxford Kelvinox-400 dilution refrigerator, with the minimum temperature
of 50 mK. The associated electrical lines were heavily filtered with
RC filters at room temperature with a cutoff frequency around 100Hz
and with thermo coaxial cable filters at low temperature with a cutoff
frequency around 1 GHz. (2) Quantum Design Physical Properties Measurement
System equipped with a He-3 insert, with no filtering system other
than having the electrical lines twisted in pairs.

To bypass effects of noise, the time domain data was taken only in
the refrigerator 1. The electrical measurements were performed with
a bandwidth of $12$ Hz and a repetition rate of $3$ Hz. At a fixed
current in the transition regime, voltage was continuously measured
for 1800 seconds, exhibiting random switching in real time. To extract
characteristic lifetimes of the superconducting $\tau_{sc}$ and the
voltage plateau $\tau_{vp}$ states the first 100-seconds of data
was discarded, in order to bypass possible transients. After that,
whenever the voltage crossed above or below a threshold value (defined
as 3 times the noise floor of the measurement setup: $\sim80$ nV),
the switching into or out of the voltage plateau state was recorded.
The time intervals between two consecutive switching events defined
residence times in one state. Finally, $\tau_{sc}$ and $\tau_{vp}$,
reported in the main text, are mean values of the stochastic residence
time sequence, collected over 1800 seconds.

\section{Acknowledgments}

Experimental work at Minnesota was supported by the DOE Office of
Basic Energy Sciences under Grant No. DE-FG02-02ER4600. Samples were
fabricated in the Nano Fabrication Center, which receives funding
from the NSF as a part of the NNIN, and were characterized in the
Characterization Facility, University of Minnesota, a member of the
NSF-funded Materials Research Facilities Network (http://www.mrfn.org)
via the MRSEC program. AK was supported by DOE Contract No. DE-FG02-08ER46482.

\section{Contributions}

All the authors contribute exclusively to this work.

\section{Competing financial interests}

The authors declare no competing finacial interests.
\end{document}


\section{Supplemental experimental data}

\begin{table*}
\centering %
\begin{tabular}{ccccccccc}
\toprule
 & \multicolumn{4}{c}{\centering\textit{Growth Parameters}} & \multicolumn{4}{c}{\centering\textit{Measured Parameters}}\tabularnewline
\cmidrule(l){2-5} \cmidrule(l){6-9} Sample  & Material  & Length ($\mu$m)  & Height (nm)  & Width (nm)  & $T_{c}$ (K)  & Rn ($\Omega$)  & $V_{0}$ (V)  & $V_{0}/\bigtriangleup_{0}$ \tabularnewline
\midrule
A  & Zn  & 4.5  & 65  & 100  & 0.76  & 52  & 52.5  & 0.46\tabularnewline
B  & Zn  & 4.0  & 60  & 100  & 0.77  & 41  & 41.0  & 0.35\tabularnewline
C  & Zn  & 4.5  & 65  & 110  & 0.75  & 53  & 44.7  & 0.40\tabularnewline
D  & Al  & 3.0  & 50  & 50  & 1.25  & 23  & 93.2  & 0.49\tabularnewline
\bottomrule
\end{tabular}\caption{List of parameters for all studied samples. Critical temperature $T_{c}$
is taken as the temperature of the half-normal resistance at a low
applied current of $0.1\,\mu A$; the BCS energy gap is defined as
$2\Delta_{0}=3.52T_{c}$. Plateau voltage $V_{0}$ is the averaged
voltage over the entire plateau regime. }

\label{tab:parameters}
\end{table*}

\begin{figure}
\includegraphics{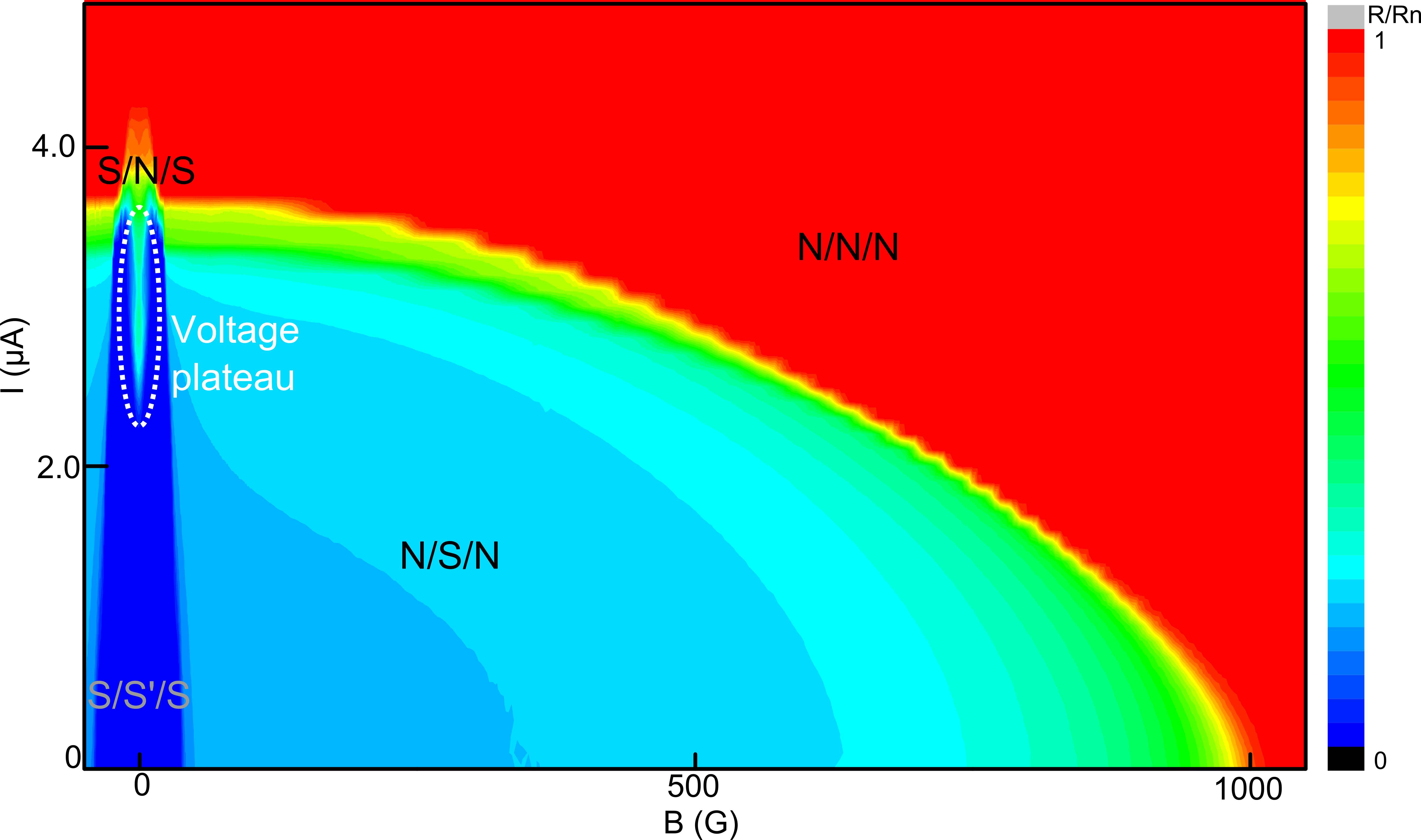}

\caption{Resistance color map of sample B, along the axes of magnetic field
and current. Regimes are labeled according to the states of the lead/nanowire/lead
system. The highlighted area is the voltage plateau regime discussed
in the main text. }

\label{phaseDiagram}
\end{figure}

Figure \ref{phaseDiagram} shows the resistance false-colored map
for sample B at $450$ mK, as a function of perpendicular magnetic
field and applied current. It complements the phase diagram of Fig.~2e
of the main text, showing a broader range of the perpendicular magnetic
fields. One may distinguish between different regimes of the total
system, as denoted in the figure: 1) N/N/N regime (red-colored): at
high fields and high currents, both the nanowire and the leads are
driven into the normal state. As a consequence, the system retains
its full normal resistance. 2) S/S'/S regime (blue-colored): at low
currents and low fields, both the nanowire and the leads are superconducting.
3) N/S/N regime (light blue-colored): at a critical field $\thicksim40$
G the leads turn normal; the narrow wire expels the field and thus
remains superconducting until much larger fields. The system exhibits
a small non-zero resistance. We believe, it is primerely coming from
the parts of the wire of length $\sim\xi$ which are in direct proximity
to the normal leads. As the field increases it weakens the superconductivity
in the wire, increasing its coherence length and thus the proximity-induced
resistance. The entire wire turns into the normal state at $\thicksim1000$
G. 4) S/N/S regime (light green-colored near zero field): at low field
and high currents, the system is in the regime, where the leads are
superconducting while the nanowire is driven into normal state by
the applied current. This is the regime that is described as the ``proximitized
normal'' in the main text. The nanowire is almost entirely normal
safe for a slightly reduced resistance, due to the proximity with
the superconducting leads. 5) The \textit{voltage plateau regime}
(encircled by white dotted line) takes place at low fields and intermediate
currents. Distinct from S/N/S regime, the voltage plateau regime is
seen to be embedded into the S/S'/S regime. This suggests that, despite
presence of a finite voltage $V_{0}$ and thus the dissipation, the
superconducting order in the nanowire is not entirely suppressed.

\begin{figure}
\includegraphics{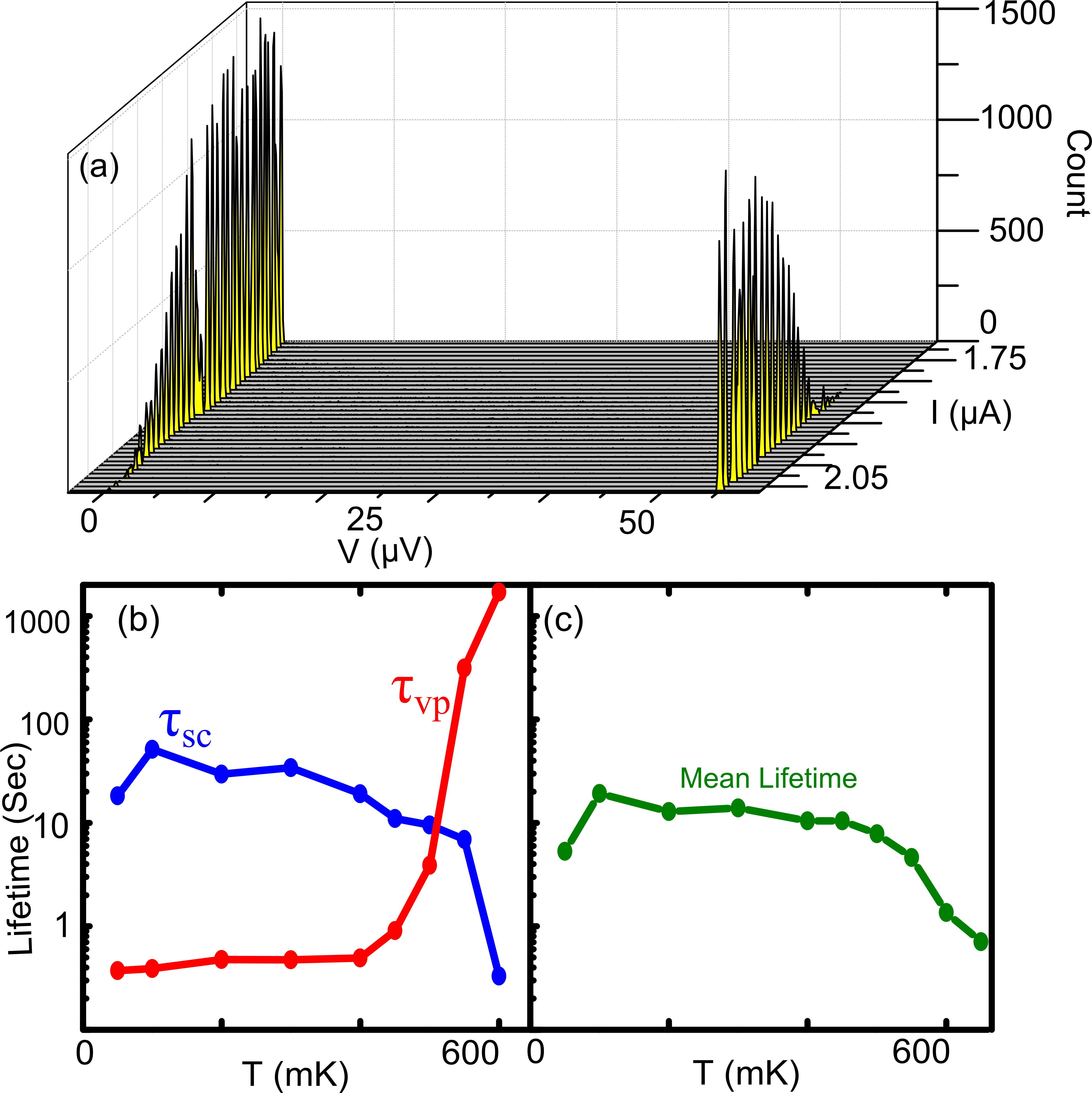}

\caption{a) Voltage distributions for various currents in the transition region.
For each current the distribution is extracted from 5400 single-shot
measurements at zero magnetic field, at 50 mK. b) $\tau_{sc}$ and
$\tau_{vp}$ as functions of temperature for current 1.95 $\mu$A.
c) The mean lifetime of the system, defined as the mean switching
time at the current where $\tau_{sc}=\tau_{vp}$, as a function of
temperature. }

\label{switchingTemp}
\end{figure}

Extensive measurements at the onset of the voltage plateau were performed
by sustaining a fixed current for 1800 seconds and repeatedly measuring
voltage with a repetition rate of 3 Hz. Total of 5400 measurements
were collected and the distribution function of measured voltages
was constructed. Figure \ref{switchingTemp}(a) shows such distribution
functions plotted for different currents across the transition regime
for sample A at 50 mK. The graph exhibits two distinct peaks: one
centered at zero voltage and the other centered at the voltage plateau.
This provides evidence to a bistable nature of the system across the
transition regime, meaning it has non-zero probabilities of occupying
the superconducting state and the voltage plateau states. Though the
mechanism that drives this switching remains unknown. Some important
insights may be obtained from measurements at different temperatures.
Shown in Fig. \ref{switchingTemp}(b), is the temperature dependence
of $\tau_{vp}$ and $\tau_{sc}$, defined in the main text, at the
current of $1.95\mbox{ \ensuremath{\mu A}}$. One can clearly see
that neither $\tau_{vp}$ nor $\tau_{sc}$ is significantly modified
by the increasing temperature up to about $\sim400$ mK. At even higher
temperature stability of the dissipative state rapidly increases,
while that of the superconducting state -- decreases. To further quantify
the temperature dependence of the switching rates we plot, Fig. \ref{switchingTemp}(c),
mean lifetime of the system at a current when $\tau_{vp}=\tau_{sc}$.
Similarly to the behavior of $\tau_{vp}$ and $\tau_{sc}$, increasing
the temperature has a negligible impact on the system below $\sim400$
mK. Above it, increasing temperature leads to the reduction of the
mean lifetime, indicating the switchings occur more frequently at
higher temperatures.

\section{Supplemental theoretical discussion}

We assume that the order parameter $\Delta(x)$ coincides locally
with the spectral gap, while the local density of states is given
by $\nu\epsilon/\sqrt{\epsilon^{2}-\Delta^{2}(x)}$. Strictly speaking,
this is not the case in the presence of a supercurrent and/or an external
magnetic field. In these cases the peak in the density of states is
rounded and the spectral gap is somewhat smaller than $\Delta$. However
since all relevant currents and fields are much less than the corresponding
depairing values, one may safely neglect this difference. For the
time being we shall also disregard the inelastic relaxation processes,
having in mind that the nanowires are shorter than the inelastic length.
Under these assumptions the Usadel equation \cite{Kamenev2011,Usadel1970}
for the distribution function takes the form of the continuity equation
for the two components of the distribution function $F(\epsilon,x,t)$,
defined in the main text
\begin{equation}
\partial_{t}F_{L/T}+\partial_{x}\big[D_{L/T}\partial_{x}F_{L/T}\big]=0\,,\label{eq:Usadel}
\end{equation}
where $D_{L}=D$ and $D_{T}(\epsilon,x)=D\epsilon^{2}/(\epsilon^{2}-\Delta^{2}(x))$
and $D$ is the normal-state diffusion constant.

Let $x_{1,2}(\epsilon)$ be the points of Andreev reflections for
quasiparticles with energy $\epsilon$ from left/right leads, i.e.
$\Delta(x_{1,2})=\epsilon$. Due to the $x$-reflection plus particle-hole
symmetries, $x\to-x$ and $\epsilon\to-\epsilon$, of the system $x_{1,2}(\epsilon)=\mp x(\epsilon\mp V/2)$.
The stationary solution of the Usadel equation (\ref{eq:Usadel})
takes the form
\begin{equation}
F(\epsilon,x)=F_{L}(\epsilon)+J_{T}(\epsilon)x\,,\label{eq:S1}
\end{equation}
where due to particle-hole symmetry, $F_{L}(\epsilon)=-F_{L}(-\epsilon)$
and $J_{T}(\epsilon)=J_{T}(-\epsilon)$. At the two Andreev interfaces
the distribution function obeys the following boundary conditions:
\begin{eqnarray}
F(\epsilon,x_{1,2})=-F(-\epsilon\pm V,x_{1,2})\,;\quad\quad\partial_{x}F(\epsilon,x_{1,2})=\partial_{x}F(-\epsilon\pm V,x_{1,2})\,.
\end{eqnarray}
Substituting Eq.~(\ref{eq:S1}), shifting the energy by $V/2$ and
employing the symmetry properties, one finds
\[
F_{L}\!\left(\epsilon\!+\!{V/2}\right)-J_{T}\!\left(\epsilon\!+\!{V/2}\right)x(\epsilon)\!=\! F_{L}\!\left(\epsilon\!-\!{V/2}\right)+J_{T}\!\left(\epsilon\!-\!{V/2}\right)x(\epsilon)
\]
and
\[
J_{T}(\epsilon+V/2)=J_{T}(\epsilon-V/2)\,.
\]
In the leading order in voltage one finds $J_{T}(\epsilon)=J_{T}=\mbox{const}$,
where
\[
J_{T}=\frac{V}{2x(\epsilon)}\,\partial_{\epsilon}F_{L}(\epsilon)\,.
\]
Bringing back slow time dependence and weak inelastic relaxation,
one may rewrite these results \cite{Nagaev2001} as the diffusive
equation in the energy direction for the longitudinal (antisymmetric)
component
\begin{equation}
\partial_{t}F_{L}(\epsilon,t)=\partial_{\epsilon}\Big({\cal D}(\epsilon)\partial_{\epsilon}F_{L}(\epsilon,t)\Big)+I_{in}[F_{L}]\,,\label{eq:S5}
\end{equation}
where the energy diffusion coefficient is given by \cite{Nagaev2001}
\begin{equation}
{\cal D}(\epsilon)=\frac{DV^{2}}{L^{2}(\epsilon)}\,;
\end{equation}
here $L(\epsilon)=2x(\epsilon)$ is the length between the two Andreev
boundaries for quasiparticles at energy $\epsilon$. The inelastic
collision term $I_{in}$ in the simplest relaxation time approximation
takes the form
\[
I_{in}[F_{L}]=\frac{1}{\tau_{in}}\left(\tanh\frac{\epsilon}{2T}-F_{L}(\epsilon,t)\right)\,.
\]
Although these equations were rigorously derived only in the limit
of small voltage, they remain at least semi-quantitatively valid all
the way up to $V\approx\Delta_{0}$.

Let us assume first that the order parameter is completely suppressed
in the middle of the wire. This is the case at the {\em top} threshold
current current $I_{t}$. In the absence of the inelastic relaxation
the corresponding stationary solution of Eq.~(\ref{eq:S5}) at $T=0$
takes the form $F_{L}(\epsilon)=\epsilon/\tilde{\Delta}$ for $|\epsilon|<\tilde{\Delta}$
and $F_{L}(\epsilon)=\mbox{sign}(\epsilon)$ for $|\epsilon|>\tilde{\Delta}$.
Here $\tilde{\Delta}=\Delta_{0}+V_{t}/2$ is the energy window, where
multiple Andreev scattering take place and we put $L(\epsilon)=L$.
Substituting it into the self-consistency relation (2) of the main
text with $\Delta(x)=0$, one obtains $\ln(2\tilde{\Delta}/\Delta_{0})=1$,
which results in $V_{t}=(e-2)\Delta_{0}\approx0.72\Delta_{0}$.

Inelastic relaxation cools down the quasiparticles, leading to a somewhat
larger $V_{t}$. Indeed, assuming for simplicity energy independent
$\tau_{in}$, one finds for the stationary solution of Eq.~(\ref{eq:S5})
for e.g. $\epsilon>0$
\[
F_{L}(\epsilon)=\frac{e^{2\tilde{\Delta}/\epsilon_{in}}(1-e^{-\epsilon/\epsilon_{in}})-(1-e^{\epsilon/\epsilon_{in}})}{e^{2\tilde{\Delta}/\epsilon_{in}}-1}\,,
\]
where $\epsilon_{in}=\sqrt{V^{2}D/L^{2}\tau_{in}}=VL_{in}/L$. Substituting
into the self-consistency relation, one finds to the leading order
in $\tilde{\Delta}/\epsilon_{in}<1$: $\ln(2\tilde{\Delta}/\Delta_{0})=1+\tilde{\Delta}^{2}/12\epsilon_{in}^{2}$.
This may be satisfied only if $\epsilon_{in}>0.9\Delta_{0}$, while
for a faster inelastic relaxation the multiple Andreev reflections
can't suppress the superconductivity. For a slower inelastic relaxation
the top voltage (the one the system acquires at a current slightly
larger than the top threshold current, i.e. $I\gtrsim I_{t}$) is
found to be $V_{t}\approx\Delta_{0}(0.72+0.81(L/L_{in})^{2})$.

We turn now to the bottom critical current $I_{b}$. In this case
the order parameter is only partially suppressed down to $0<\Delta<\Delta_{0}$
inside the wire. This requires the voltage $V_{0}\approx2\Delta$
to cut through the order parameter and allow for PSs. In other words:
frequency of PS's is given by $2V$, while duration of a single PS
is about $4\Delta$ \cite{Arutyunov2008}; for the order parameter
to be continuously suppressed at PS location the voltage should be
at least $2\Delta$. The continious train of PSs equilibrates quasiparticles
and holes immediately below and above the suppressed gap, leading
to the boundary condition $F_{L}(|\epsilon|<\Delta)=0$. The corresponding
solution of Eq.~(\ref{eq:S5}) with no inelastic relaxation takes
the form
\begin{equation}
F_{L}(\epsilon)=\left\{ \begin{array}{ll}
0 & 0<\epsilon<\Delta\,;\\
(\epsilon-\Delta)/(\Delta_{lead}-\Delta)\quad & \Delta<\epsilon<\Delta_{lead}\,;\\
1 & \Delta_{lead}<\epsilon\,.
\end{array}\right.\label{eq:distribution}
\end{equation}
It is depicted in Fig. 3 of the main text for the case $\Delta_{lead}=\Delta_{0}$,
i.e. at $T=0$ and in the absence of magnetic field. Substituting
this distribution function into the self-consistency relation, one
finds a closed transcendental equation for $\delta=\Delta/\Delta_{lead}$:
\begin{equation}
\sqrt{1-\delta^{2}}-\ln\left(1+\sqrt{1-\delta^{2}}\right)+(1-\delta)\ln(\Delta_{0}/\Delta_{lead})=\delta\ln(1/\delta)\,.\label{eq:trans}
\end{equation}
For $\Delta_{lead}=\Delta_{0}$ it admits two solutions: $\delta=1$
-- the superconducting state with no PSs and $\delta=0.17$ -- the
dissipative state with the PSs train. Therefore in the presence of
phase slips and in the absence of inelastic relaxation the wire sustains
the (nearly space-independent) suppressed order parameter $\Delta\approx0.17\Delta_{0}$.
As before, the inelastic relaxation cools down the quasiparticles
and thus increases the suppressed self-consistent order parameter
of the wire. The calculation analogous to the one outlined above leads
to $\Delta\approx\Delta_{0}(0.17+0.53(L/L_{in})^{2})$. As a result
the corresponding voltage plateau acquires the value $V_{0}\approx\Delta_{0}(0.34+1.06(L/L_{in})^{2})$.

Voltage causes a normal current $I_{n}=\nu\!\int\! d\epsilon\, D_{T}(\epsilon)J_{T}(\epsilon)$
to flow across the wire. For not too large voltage $V\lesssim2\Delta_{0}$
one finds $J_{T}(\epsilon)=\partial_{\epsilon}F_{L}(\epsilon)eV/L$.
The energy integration over the range $|\epsilon|<\Delta_{0}+V/2$
yields the logarithmically divergent integral, which is to be cut
at $|\epsilon-\Delta|\approx V$. As a result, for $V=V_{0}$ one
obtains $I_{n}=(1+2\delta-\delta\ln\delta)V_{0}/R_{n}$, where $R_{n}=L/(e^{2}\nu D)$
is the normal state resistance of the wire. We thus find that for
the dissipative state to occur a minimal current $I_{b}\approx1.64(V_{0}/R_{n})$
should be supplied to pass through the wire. Since the order parameter
is totally suppressed only in PS region of size $\sim\xi$, the wire
may still support the supercurrent $I_{sc}$ in addition to the normal
one. While the latter is carried by quasiparticles undergoing multiple
Andreev reflections, the former is due to the small phase gradient
of the (suppressed) order parameter $\Delta(x)=\Delta e^{i\phi x/L}$.
As a result the current in excess of $I_{b}$ is carried as the supercurrent,
exhibiting the voltage plateau pattern. However once the total current
riches $I_{t}$, the system becomes unstable against transition to
the state with zero order parameter across the wire, discussed above.
This eliminates the supercurrent and thus for $I>I_{t}$ the wire's
resistance is close to $R_{n}$.

The main effect of the magnetic field or elevated temperature is to
suppress the equilibrium order parameter inside the leads $\Delta_{lead}<\Delta_{0}$.
This narrows the energy window for the Andreev diffusion, Eq.~(\ref{eq:S5}).
Numerical investigation of transcendental equation (\ref{eq:trans})
shows that the non-trivial solution is only possible as long as $\Delta_{lead}\geq0.78\Delta_{0}$,
otherwise the dissipative state with a non-zero order parameter is
not sustainable. The quantitative comparison in case of the perpendicular
magnetic field requires knowledge of the leads order parameter next
to the contact with the wire. This would demand knowing precise location
of the vortices in the leads. On the other hand, the temperature dependence
is rather straightforward. Taking as an example sample A with the
critical temperature $T_{c}=0.76$ K and using standard BCS expressions
one estimates that the order parameter is suppressed by the factor
0.78 relative to its $T=0$ value at $T\approx0.75T_{c}\approx0.59$
K. This is in a reasonable agreement with $T=0.65$ K observed as
the upper temperature limit of the voltage plateau, Fig 1b of the
main text. The difference may be again attributed to the residual
inelastic processes, which cool down the non-equilibrium distribution.